\documentclass[10pt,conference]{IEEEtran}

\usepackage{research5}
\usepackage{multicol}

\newtheorem{dfn}{Definition}
\newtheorem{thm}{Theorem}
\newtheorem{cor}{Corollary}

\begin{document}

\title{Sending a Bi-Variate Gaussian Source over a Gaussian MAC}

\author{\authorblockN{Amos Lapidoth ~ ~ ~ Stephan Tinguely}
\authorblockA{Signal and Information Processing Laboratory\\
Swiss Federal Institute of Technology (ETH) Zurich, Switzerland\\
\texttt{ \{lapidoth, tinguely\}@isi.ee.ethz.ch}}}

\maketitle

\begin{abstract}
  We consider a problem where a memoryless bi-variate Gaussian source
  is to be transmitted over an additive white Gaussian multiple-access
  channel with two transmitting terminals and one receiving
  terminal. The first transmitter only sees the first source component
  and the second transmitter only sees the second source component. We
  are interested in the pair of mean squared-error distortions at
  which the receiving terminal can reproduce each of the source
  components.
  
  It is demonstrated that in the symmetric case, below a certain
  signal-to-noise ratio (SNR) threshold, which is determined by the
  source correlation, uncoded communication is optimal. For SNRs above
  this threshold we present outer and inner bounds on the achievable
  distortions. 
\end{abstract}

\section{Introduction}
We consider the situation where a memoryless bi-variate Gaussian
source is to be transmitted over an additive white Gaussian
multiple-access channel with two transmitting terminals and one
receiving terminal. Each of the two source components is fed to a
different average-power constrained encoder. Our interest lies in
the achievable expected squared-error distortion region. We show that
in the symmetric case, where the source components are of the same
variance and the transmitting terminals are subjected to
the same average power constraint, uncoded transmission is optimal
below a threshold signal-to-noise ratio (SNR) that is determined by
the correlation between the source components. For SNRs above this
threshold we provide outer and inner bounds on the achievable
distortions. 

The problem at hand can be viewed as the Gaussian version of the
problem addressed by Cover, El Gamal and Salehi
\cite{cover_elgamal_salehi80} (see also \cite{dueck81} and
\cite{kang_ulukus05}). It also appears to be closely related to the
quadratic Gaussian CEO problem \cite{berger_vishwanathan97, oohama98}
and the quadratic Gaussian two-terminal source-coding problem
\cite{wagner_tavildar_vishwanath05, oohama97}. However, it differs in
character from the CEO problem and from the two-terminal source coding
problem in that no error-free bit-pipes of finite rates can be
assumed. This is due to the fact that the source-channel separation
theorem does not apply to our situation. Furthermore, the CEO problem
focuses on the reconstruction of a single Gaussian random variable,
whereas in our case the interest lies in the reconstruction of both
source components.

\section{Problem Statement}

The time-$k$ output $Y_{k} \in \Reals$ of the discrete-time two-user
additive white Gaussian multiple-access channel is given by
\begin{displaymath}
Y_k = x_{1,k} + x_{2,k} + Z_k,
\end{displaymath}
where $x_{1,k} \in \Reals$ denotes the time-$k$ symbol transmitted by
the first transmitter, $x_{2,k} \in \Reals$ is the time-$k$ symbol
transmitted by the second transmitter, and $Z_{k}$ denotes the
time-$k$ noise term. The noise terms $\{Z_{k}\}$ are independent
identically distributed (IID) zero-mean variance-$N$ Gaussian random
variables that are independent of the input sequences $(\{x_{1,k}\},
\{x_{2,k}\})$. We shall consider the case where Transmitter~1 and
Transmitter~2 are average-power limited to $P_{1}$ and $P_{2}$
respectively. See (\ref{eq:power}) ahead.

At time $k$ the source emits the pair $(S_{1,k}, S_{2,k})$ where the
$\{ (S_{1,k}, S_{2,k}) \}$ are IID zero-mean Gaussians of covariance
\begin{displaymath}
\cov{S} = \left( \begin{array}{c c}
\sigma_1^2 & \rho \sigma_1 \sigma_2\\
\rho \sigma_1 \sigma_2 & \sigma_2^2
\end{array} \right),
\end{displaymath}
with $\rho \in [-1,1]$, and $0 < \sigma_i^2 < \infty$, $i = 1,2$. 

The sequence $\{S_{1,k}\}$ is fed to Transmitter~1 and the sequence
$\{S_{2,k}\}$ is fed to Transmitter~2. Based on the channel output we
wish to reconstruct the source vector. The performance criterion we
focus on is the expected squared-error distortions in reconstructing
each of the components of the source vector.

\begin{dfn}
Given $\sigma_{1}, \sigma_{2} > 0$, $\rho \in [-1,1]$,
and $P_{1}, P_{2} > 0$ we say that the tuple $\bigl(D_{1}, D_{2},
\sigma^{2}_{1}, \sigma_{2}^{2}, \rho, P_{1}, P_{2} \bigr)$ is
achievable if there exists a sequence of encoder pairs $(f_{1}^{(n)},
f_{2}^{(n)})$ 
\begin{equation*}
  f_{i}^{(n)} : \Reals^{n} \rightarrow \Reals^{n}, \quad i=1,2
\end{equation*}
and a sequence of reconstruction pairs $(\phi_{1}^{(n)},
\phi_{2}^{(n)})$ 
\begin{equation*}
  \phi_{i}^{(n)} : \Reals^{n} \rightarrow \Reals^{n}, \quad i=1,2
\end{equation*}
such that the average power constraints are satisfied
\begin{equation}
  \label{eq:power}
  \frac{1}{n} \E{\|f_{i}^{(n)}\bigl(S_{i}^{n}\bigr)\|^{2}} \leq
  P_{i}, \quad i=1,2
\end{equation}
and
\begin{multline}
  \varlimsup_{n \rightarrow \infty} \frac{1}{n} \textsf{E} \left[ \Big\| (S_{i,1},
      \ldots, S_{i,n}) \right. \\
\left. \left. - \phi_{i}^{(n)}\left( f_{1}^{(n)}(S_{1}^{n}) +
     f_{2}^{(n)}(S_{2}^{n}) + (Z_{1}, \ldots, Z_{n})
   \right) \right\|^{2} \right] \leq D_{i},\\ \quad i=1,2,
\end{multline}
whenever $\{(S_{1,k},S_{2,k})\}$ are IID zero-mean bi-variate Gaussian
vectors of covariance matrix $\cov{S}$ as above and $\{Z_{k}\}$ are
IID zero-mean variance-$N$ random variables that are independent of
$\{(S_{1,k},S_{2,k})\}$. Here we used the shorthand notation where
$S_{1}^{n}$ denotes $(S_{1,1}, \ldots, S_{1,n})$ and similarly for
$S_{2}^{n}$.
\end{dfn}

The problem we address here is, for given
$\sigma_{1}^{2}, \sigma_{2}^{2}, \rho, P_{1}, P_{2}$, to find the set
of pairs $(D_{1}, D_{2})$ such that $(D_{1}, D_{2}, \sigma_{1}^{2},
\sigma_{2}^{2}, \rho, P_{1}, P_{2})$ is achievable.

By the symmetric version of this problem we shall refer to the case
where $\sigma_{1}^{2} = \sigma_{2}^{2}$, where $P_{1} = P_{2}$, and
where we seek the set of pairs $(D,D)$ that are achievable. That is,
if we set $\sigma^{2} =  \sigma_{1}^{2} = \sigma_{2}^{2}$ and $P =
P_{1} = P_{2}$ then we are interested in 
\begin{multline}
 D^{*}(P,N,\sigma^{2}, \rho) \triangleq \sup \{D: (D, D, \sigma^{2},
\sigma^{2}, \rho, P, P)\\ \text{is achievable}\}.
\end{multline}

\section{Preliminary Remarks}
Before discussing our results, we make three remarks regarding the
general nature of the problem. The firs two remarks show that there is
no loss in generality by assuming that the correlation coefficient is
non-negative and that the source components are of equal variance. As
a consequence we shall assume for the remainder that $\sigma_1^2 =
\sigma_2^2 = \sigma^2$ and that $\rho \in [0,1]$. The third remark
addresses a convexification issue of the distortion regions.

\begin{enumerate}
\item The optimal distortion region depends on the correlation
  coefficient only via its absolute value $|\rho |$. That is, the
  tuple $(D_{1}, D_{2}, \sigma_{1}^{2}, \sigma_{2}^{2}, \rho, P_{1},
  P_{2})$ is achievable if, and only if, the tuple $(D_{1}, D_{2},
  \sigma_{1}^{2}, \sigma_{2}^{2}, -\rho, P_{1}, P_{2})$ is
  achievable.
  
  To see this note that if $(f_{1}^{(n)}, f_{2}^{(n)}, \phi_{1}^{(n)},
  \phi_{2}^{(n)})$ achieves the distortion $(D_{1}, D_{2})$ for the
  source of correlation coefficient $\rho$, then $(\tilde{f}_{1}^{(n)},
  f_{2}^{(n)}, \tilde{\phi}_{1}^{(n)}, \phi_{2}^{(n)})$ where
  \begin{equation*}
    \tilde{f}_{1}^{(n)}(S_{1}^{n}) = f_{1}^{(n)}(-S_{1}^{n})
  \end{equation*}
  and 
  \begin{equation*}
    \tilde{\phi}_{1}^{(n)}(Y_{1},\ldots, Y_{n}) = -
    \phi_{1}^{(n)}(Y_{1},\ldots, Y_{n}) 
  \end{equation*}
  achieves  $(D_{1}, D_{2})$ on the source with correlation
  coefficient $-\rho$.
  
\item The optimal distortions scale linearly with the source
  variances. That is, if $\alpha_{1}, \alpha_{2}$ are positive then
  $(D_{1}, D_{2}, \sigma_{1}^{2}, \sigma_{2}^{2}, \rho, P_{1}, P_{2})$
  is achievable if, and only if, $(\alpha_{1}^{2}D_{1},
  \alpha_{2}^{2}D_{2}, \alpha_{1}^{2} \sigma_{1}^{2}, \alpha_{2}^{2}
  \sigma_{2}^{2}, \rho, P_{1}, P_{2})$ is achievable. Consequently,
  there is a simple linear transformation from the set of tuples
  $(D_{1}, D_{2})$ for which $(D_{1}, D_{2}, \sigma_{1}^{2},
  \sigma_{2}^{2}, \rho, P_{1}, P_{2})$ is achievable and the set of
  tuples $(\tilde{D}_{1}, \tilde{D}_{2})$ for which $(\tilde{D}_{1},
  \tilde{D}_{2}, \alpha_{1}^{2} \sigma_{1}^{2}, \alpha_{2}^{2}
  \sigma_{2}^{2}, \rho, P_{1}, P_{2})$ is achievable.

  To see this note that if $(f_{1}^{(n)}, f_{2}^{(n)}, \phi_{1}^{(n)},
  \phi_{2}^{(n)})$ demonstrate the achievability of $(D_{1}, D_{2},
  \sigma_{1}^{2}, \sigma_{2}^{2}, \rho, P_{1}, P_{2})$ then the
  encoders
  \begin{equation*}
    \tilde{f}_{i}^{(n)}(S_{i}^{n}) = f_{i}^{(n)}(S_{i}^{n}/\alpha_{i})
    \quad i=1,2
  \end{equation*}
  and the reconstructions
  \begin{equation*}
    \tilde{\phi}_{i}^{(n)}(Y_{1}, \ldots, Y_{n}) = \alpha_{i} \cdot
    \phi_{i}^{(n)}(Y_{1}, \ldots, Y_{n}), \quad i=1,2
  \end{equation*}
  demonstrate the achievability of the tuple $(\alpha_1^2 D_{1}, \alpha_2^2
  D_{2}, \alpha_{1}^{2} \sigma_{1}^{2}, \alpha_{2}^{2} \sigma_{2}^{2},
  \rho, P_{1}, P_{2})$.
  
  Applying the same argument in the other direction with scalings by
  $1/\alpha_{1}$ and $1/\alpha_{2}$ concludes the proof.

\item The achievable distortion is a convex function of the
  power constraints $(P_1,P_2)$. That is, if $(D_{1}, D_{2}, 
  \sigma_{1}^{2}, \sigma_{2}^{2}, \rho, P_{1}, P_{2})$ and $(\tilde{D}_{1}, \tilde{D}_{2}, 
  \sigma_{1}^{2}, \sigma_{2}^{2}, \rho, \tilde{P}_{1}, \tilde{P}_{2})$
  are achievable then 
  \begin{equation*}
    \bigl(\lambda D_{1} + \bar{\lambda} \tilde{D}_{1}, \lambda D_{2} + 
    \bar{\lambda}\tilde{D}_{2},  \sigma_{1}^{2}, \sigma_{2}^{2}, \rho,
    \lambda P_{1} + \bar{\lambda} \tilde{P}_{1}, \lambda P_{2} + 
    \bar{\lambda}\tilde{P}_{2}) \bigr)
  \end{equation*}
  is achievable for any $\lambda \in [0,1]$, where $\bar{\lambda} =
  (1-\lambda)$.

  This follows by a simple time-sharing argument
\end{enumerate}

\section{Main Results}
We present necessary conditions as well as sufficient conditions for
achievability. In certain cases they agree. The proofs of those
conditions will be discussed in the next section.

Our first result is a necessary condition for the achievability of
$(D_1, D_2, \sigma_1^2 , \sigma_2^2, \rho, P_1, P_2)$.

\begin{thm}
\label{thm:converse}
A necessary condition for the achievability of $(D_1, D_2, \sigma^2,
\sigma^2, \rho, P_1, P_2)$ is that
\begin{displaymath}
\frac{1}{2} \log \left( 1 + \frac{P_1 + P_2 + 2 \rho \sqrt{P_1
      P_2}}{N} \right) \geq R(D_1,D_2),
\end{displaymath}
where the expression for $R(D_1,D_2)$ varies, depending on the values
of $(D_1,D_2)$. There are three cases. If $(D_1,D_2)$ are in the set
\begin{displaymath}
\left\{ D_1 \leq \sigma^2 (1-\rho), D_2 \leq (\sigma^2(1-\rho^2) -
  D_1) \frac{\sigma^2}{\sigma^2-D_1} \right\},
\end{displaymath}
then
\begin{displaymath}
R(D_1,D_2) = \frac{1}{2} \log_2 \left(
  \frac{\sigma^4 (1-\rho^2)}{D_1 D_2} \right).
\end{displaymath}
If $(D_1,D_2)$ are in the set
\begin{multline*}
\Big\{ 0 \leq D_1 \leq \sigma^2,\\
(\sigma^2(1-\rho^2) - D_1) \frac{\sigma^2}{\sigma^2-D_1} \leq D_2
\leq \sigma^2 (1-\rho^2) + \rho^2 D_1 \Big\},
\end{multline*}
then
\begin{multline*}
R(D_1,D_2) =\\
\frac{1}{2} \log_2 \left(
  \frac{\sigma^4 (1-\rho^2)}{D_1D_2 - \left( \rho \sigma^2 -
      \sqrt{(\sigma^2-D_1)(\sigma^2-D_2)} \right)^2} \right),
\end{multline*}
and if $(D_1,D_2)$ are in the set
\begin{displaymath}
\Big\{ 0 \leq D_1 \leq \sigma^2, D_2 > \sigma^2 (1-\rho^2) + \rho^2 D_1 \Big\} . 
\end{displaymath}
then
\begin{displaymath}
R(D_1,D_2) = \frac{1}{2} \log_2 \left(
  \frac{\sigma^2}{D_1} \right).
\end{displaymath}
\end{thm}

\begin{cor}\label{cor:1}
In the symmetric case where $P_1 = P_2$, we obtain
\begin{displaymath}
D^{\ast}(\sigma^2, \rho, P, N) \geq \left\{ \begin{array}{l l}
\sigma^2 \frac{P (1-\rho^2) +N}{2P(1 + \rho) +N} & \text{for
}\frac{P}{N} \in \left( 0, \frac{\rho}{1-\rho^2}\right]\\[5mm]
\sigma^2 \sqrt{\frac{(1-\rho^2)N}{2P(1+\rho) +N}} & \text{for }
\frac{P}{N} > \frac{\rho}{1-\rho^2}.
\end{array} \right.
\end{displaymath}
\end{cor}
\noindent
{\bf Note:} Theorem~\ref{thm:converse} can be easily extended to a much
wider class of sources and distortions. Indeed, if the source is any
memoryless bi-variate source (not necessarly zero-mean Gaussian) and
if the fidelity measures $d_{1}(s_{1}, \hat{s}_{1}), d_{2}(s_{2},
\hat{s}_{2}) \geq 0$ that are used to measure the distortion in
reconstructing each of the source components are arbitrary, then the
pair $(D_{1}, D_{2})$ is achievable with powers $P_{1}, P_{2}$ only
if 
\begin{align}
\min_{P_{\widehat{S}_1, \widehat{S}_2 | S_1, S_2}} I &(S_1,S_2;
\widehat{S}_1, \widehat{S}_2) \label{eq:RD} \\[2mm]
\text{such that } &\E{(S_1 -\widehat{S}_1)^2} \leq D_1, \nonumber \\
&\E{(S_2 -\widehat{S}_2)^2} \leq D_2, \nonumber
\end{align}
does not exceed
\begin{displaymath}
\frac{1}{2} \log \left( 1 + \frac{P_1 + P_2 + 2\rho_{\textnormal{max}}
    \sqrt{P_1 P_2}}{N} \right),
\end{displaymath}
where $\rho_{\textnormal{max}}$ is the Hirschfeld-Gebelein-R\'{e}nyi
maximal correlation between $S_{1}$ and $S_{2}$:
\begin{equation}
  \rho_{\textnormal{max}} = \sup \E{g(S_{1}) h(S_{2})}
\end{equation}
where the supremum is over all functions $g,h$ under which
\begin{equation}
 \E{g(S_{1})} = \E{h(S_{2})} = 0 \qquad \E{g^{2}(S_{1})} = \E{h^{2}(S_{2})} = 1
\end{equation}

We next present two sufficient conditions for the achievability of
$(D_1, D_2, \sigma^2, \sigma^2, \rho, P_1, P_2)$. The first is
obtained by analyzing uncoded transmission.
\begin{thm}
\label{thm:uncoded}
For $(D_1, D_2, \sigma^2, \sigma^2, \rho, P_1, P_2)$ to be achievable
it suffices that both of the following conditions hold:
\begin{align*}
D_1 &\geq \sigma^2 \left( \frac{2P_1 + 4\rho \sqrt{P_1 P_2} +
  (1+\rho^2)P_2 + N -2\sqrt{P_1 + 2\rho \sqrt{P_1 P_2} + \rho^2 P_2}
  (\sqrt{P_1}+ \rho \sqrt{P_2})}{P_1 + P_2 + \rho \sqrt{P_1 P_2} + N} \right)\\[5mm]
D_2 &\geq \sigma^2 \left( \frac{2P_2 + 4\rho \sqrt{P_1 P_2} +
  (1+\rho^2)P_1 + N -2\sqrt{P_2 + 2\rho \sqrt{P_1 P_2} + \rho^2 P_1}
  (\sqrt{P_2}+ \rho \sqrt{P_1})}{P_1 + P_2 + \rho \sqrt{P_1 P_2} +
  N} \right).
\end{align*}
\end{thm}

\begin{cor}\label{cor:2}
In the symmetric case
\begin{displaymath}
D^{\ast}(\sigma^2, \rho, P, N) \leq \sigma^2 \frac{P(1-\rho^2)
  +N}{2P(1+\rho) +N}
\end{displaymath}
\end{cor}
Combining Corollary \ref{cor:1} and Corollary \ref{cor:2}, we obtain:
\begin{cor}\label{cor:3}
For the symmetric case,
\begin{displaymath}
D^{\ast}(\sigma^2, \rho, P, N) = \sigma^2 \frac{P(1-\rho^2)
  +N}{2P(1+\rho) +N}, \qquad \text{if } \frac{P}{N} < \frac{\rho}{1-\rho^2}
\end{displaymath}
i.e., uncoded transmission is optimal for all $P/N < \rho/(1-\rho^2)$.
\end{cor}

The second sufficient condition follows from analyzing the scheme
where the encoding functions $f_i^{(n)}(s_i^n)$, $i=1,2$, are randomly
generated independent rate-$R_i$ vector quantizers, i.e.~the channel
inputs are the rate-$R_i$ vector quantized source sequences.
\begin{thm}
\label{thm:coded}
The tuple $(D_1, D_2, \sigma^2, \sigma^2, \rho, P_1, P_2)$ is
achievable whenever there exist rates $R_1 > 0$ and $R_2 > 0$ such
that all of the following hold:
\begin{align*}
R_1 &< \frac{1}{2} \log_2 \left(
  \frac{P_1(1-\tilde{\rho}^2)+N}{N(1-\tilde{\rho}^2)} \right)\\[3mm]
R_2 &< \frac{1}{2} \log_2 \left(
  \frac{P_2(1-\tilde{\rho}^2)+N}{N(1-\tilde{\rho}^2)} \right)\\[3mm]
R_1 + R_2 &< \frac{1}{2} \log_2 \left( \frac{P_1+P_2
    +2\tilde{\rho}\sqrt{P_1P_2}+N}{N(1-\tilde{\rho}^2)} \right)\\[3mm]
D_1 &> \sigma^2 2^{-2R_1} \cdot \frac{1-
  \rho^2(1-2^{-2R_2})}{1-\tilde{\rho}^2}\\
D_2 &> \sigma^2 2^{-2R_2} \cdot \frac{1-
  \rho^2(1-2^{-2R_1})}{1-\tilde{\rho}^2}.
\end{align*}
where $\tilde{\rho} = \rho \sqrt{(1-2^{-2R_1})(1-2^{-2R_2})}$.
\end{thm}

\begin{cor}\label{cor:4}
In the symmetric case $(D, D, \sigma^2, \sigma^2, \rho, P, P)$ is
achievable if there exists some $R > 0$ satisfying
\begin{align}
R &< \frac{1}{4} \log_2 \left( \frac{2P (1+\rho(1-2^{-2R})) +N}{N(1-
    \rho^2(1-2^{-2R})^2)} \right)\\[3mm]
D &> \sigma^2 2^{-2R} \cdot \frac{1-
  \rho^2(1-2^{-2R})}{1-\rho^2(1-2^{-2R})^2}. \label{eq:dist-vect-sym}
\end{align}
\end{cor}
Here the RHS of (\ref{eq:dist-vect-sym}) is monotonically decreasing
in $R$. Evaluating Corollary \ref{cor:4} and Corollary \ref{cor:1} for
$P/N \rightarrow \infty$ we get:
\begin{cor}\label{cor:5}
In the symmetric case
\begin{displaymath}
\lim_{P/N \rightarrow \infty} \sqrt{\frac{P}{N}} D^{*}(\sigma^2, \rho,
P, N) = \sigma^2 \sqrt{\frac{1-\rho}{2}}.
\end{displaymath}
\end{cor}

We conclude this section with a note on the superposition of the two
discussed coding schemes.
\vspace{3.2cm}

{\bf Note:} We have analyzed two coding schemes; uncoded transmission
and transmission of vector-quantized source sequences. The
superposition of those two schemes, analogous to the scheme discussed
for the single-user case in \cite{bross_lapidoth_tinguely06}, seems to
yield strict improvements of the above discussed achievable $(D_1,
D_2, \sigma^2, \sigma^2, \rho, P_1, P_2)$. Detailed results are to follow.

\section{Notes on the Derivations}

In this section we shall try to sketch the ideas behind the proofs of
the main results.
%%Space limitations preclude us from providing complete proofs.

The proof of Theorem~\ref{thm:converse} consists on one hand of
upper bounding the mutual information between the the source vectors
and the reconstructions, and on the other hand evaluating the rate
distortion function for a bi-variate Gaussian source.
The key to upper bounding the mutual information between source and
reconstructions is to use the average power constraints
\eqref{eq:power} and the limited correlation between the source
components to obtain the upper bound
\begin{equation}
  \frac{1}{n} \sum_{k=1}^{n} \Var{X_{1,k}\bigl(S_{1}^{n}\bigr) +
  X_{2,k}\bigl(S_{2}^{n}\bigr)} \leq P_{1} + P_{2} + 2 \rho \sqrt{P_{1}P_{2}}
\end{equation}
where $X_{1,k}(S_{1}^{n})$ is the $k$-th component of
$f_{1}^{(n)}(S_{1}^{n})$ and where $X_{2,k}(S_{2}^{n})$ is analogously
defined. Once this bound is established for all encoders
$f_{1}^{(n)}$, $f_{2}^{(n)}$ satisfying the power constraints \eqref{eq:power},
one can derived necessary conditions for achievability by using the
data processing inequality to upper bound the mutual information
between the source vectors and their reconstructions by the mutual
information between the transmitted waveforms and the received
waveform. This latter mutual information is upper bounded by the
capacity of the additive Gaussian noise channel subject to the power
constraint $P_{1} + P_{2} + 2 \rho \sqrt{P_{1}P_{2}}$.

The rate distortion function is obtained from evaluating
\eqref{eq:RD} under the given distortion constraints and for the given
source law $P_{S_1,S_2}$. From the maximum mutual information theorem
it follows that this minimum is achieved if and only if $S_1, S_2,
\widehat{S}_1, \widehat{S}_2$ are jointly Gaussian. The minimization
problem is then reduced to a minimization over the set of covariance
matrices of $S_1, S_2, \widehat{S}_1, \widehat{S}_2$ that satisfy the
distortion constraints and where the submatrix in $S_1, S_2$ is the
covariance matrix of the source. The minimizing covariance matrix can
be found by noticing that every relevant distortion pair can be
achieved, with minimal necessary rate, by combining a scaling of the
source with reverse waterfilling. Let $\mathcal{D}(R)$ be the set of
all distortion pairs $(d_1,d_2)$ that can be achieved on the source pair
$(S_1,S_2)$ with rate $R$, and let $\mathcal{D}_c(R)$ be the set of
$(d_1,d_2)$ that can be achieved with rate $R$ on the scaled source
$(S_1,cS_2)$. The region $\mathcal{D}_c(R)$ corresponds to the region
$\mathcal{D}(R)$ scaled by a factor $c^2$ on the $S_2$-axis. Reverse
waterfilling at rate $R$ on the unitarily decorrelated pair
$(V_1,V_2)$ of $(S_1,cS_2)$ achieves the point $(d_1^{\ast},
d_2^{\ast}) \in \mathcal{D}_c(R)$ of minimal sum $d_1 + d_2$. And
since $R$ is the minimal rate needed to achieve $(d_1^{\ast},
d_2^{\ast})$ on $(S_1,cS_2)$, and
\begin{multline*}
\min_{\substack{P_{\widehat{S}_1, \widehat{S}_2 | S_1,S_2}:\\
\E{(S_1 - \widehat{S}_1)^2} \leq d_1 \\
\E{(S_2 - \widehat{S}_2)^2} \leq \frac{1}{c^2}d_2}} I(S_1,S_2 ;
\widehat{S}_1, \widehat{S}_2 )
=\\
\min_{\substack{P_{\widehat{S}_1, \widehat{S}_2 | S_1,S_2}:\\
\E{(S_1 - \widehat{S}_1)^2} \leq d_1 \\
\E{(cS_2 - c\widehat{S}_2)^2} \leq d_2}} I(S_1,cS_2 ; \widehat{S}_1 ,
c\widehat{S}_2 ),
\end{multline*}
the rate $R$ is also the minimal rate needed to achieve
$(d_1^{\ast}, d_2^{\ast}/c^2)$ on $(S_1,S_2)$. Hence, by choosing
the appropriate scaling $c$, we can get any relevant point on the
boundary of $\mathcal{D}(R)$. The covariance matrix of
$(S_1,S_2,\widehat{S}_1,\widehat{S}_2)$ that achieves $(d_1^{\ast},
d_2^{\ast}/c^2)$ now follows from the covariance matrix of $(V_1,
V_2, \widehat{V}_1, \widehat{V}_2)$, where $(\widehat{V}_1,
\widehat{V}_2)$ result from reverse waterfilling at rate $R$ on
$(V_1,V_2)$.\footnote{We note that this idea generalizes to Gaussian
  sources with more than two components.}

The proof of Theorem~\ref{thm:uncoded} is straightforward. One merely
considers the uncoded scheme where
\begin{equation*}
  f_{i}^{(n)}(S_{i}^{n}) = \frac{\sqrt{P_{i}}}{\sigma} (S_{i,1},
  \ldots S_{i,n}), \quad i=1,2
\end{equation*}
and then analyzes the linear minimum mean squared-error estimators of
$S_{i,k}$ from $Y_{k}$. 

The proof of Theorem~\ref{thm:coded} involves an analysis of randomly
generated independent vector quantizers for the two components. The
proposed scheme is conceptually simple, but its analysis gets involved
by the included epsilons and deltas. For the sake of clarity and
brevity we shall omit these epsilons and deltas here.

The encoder for the $i$-th, $i=1,2$, source component is a rate-$R_i$
Gaussian vector quantizer that scales the quantized sequence to meet
the channel input power constraint. Its codebook $\mathcal{C}_i$
consists of $2^{nR_i}$ codewords that are chosen IID uniformly on the
surface of an $\Reals^n$-sphere of center at the origin and radius
$\sqrt{n \sigma^2(1-2^{-2R_i})}$. Encoder~$i$ chooses the codeword
$\vect{u}_i^{\ast}$ in the codebook $\mathcal{C}_i$ that is closest
(in Euclidean distance) to the source sequence $\vect{s}_i = (s_{i,1},
s_{i,2}, \ldots ,s_{i,n})$, and transmits its scaled version
\begin{align*}
\vect{x}_i &= \alpha_i \argmin_{\vect{u}
  \in \mathcal{C}_i} \norm{\vect{s}_i -\vect{u}}\\
&= \alpha_i \argmax_{\vect{u} \in
  \mathcal{C}_i} \inner{\vect{s}_i}{\vect{u}},
\end{align*}
where
\begin{displaymath}
\alpha_i = \sqrt{\frac{P_{i}}{\sigma^2(1-2^{-2R_i})}},
\end{displaymath}
and where $\inner{\cdot}{\cdot}$ denotes the standard inner product in
$\Reals^n$. The distance $\norm{\vect{s}_i - \vect{u}_i^{\ast}}$
between the source sequence $\vect{s}_i$ and its closest codeword
$\vect{u}_i^{\ast}$ approaches, with high probability, $\sigma^2 \cdot
2^{-2R_i}$ as the blocklength $n$ tends to infinity. It can be shown
that, for large $n$, the correlation coefficient between the chosen
codewords $\vect{U}_1^{\ast}$ and $\vect{U}_2^{\ast}$ is, with very
high probability, close to
\begin{displaymath}
\tilde{\rho} = \rho \sqrt{(1-2^{-2R_1})(1-2^{-2R_2})}.
\end{displaymath}
This coefficient $\tilde{\rho}$ plays a central role in this coding
scheme.

The decoding is performed in two parts. First the transmitted codeword
pair is recovered, and then this codeword pair is used to make linear
estimates of the source sequences. To recover the transmitted pair
$(\vect{u}_1^{\ast}, \vect{u}_2^{\ast})$, the decoder seeks, among all
``jointly typical'' pairs $(\vect{u}_{1}, \vect{u}_{2}) \in
\mathcal{C}_1 \times \mathcal{C}_2$, i.e among all pairs satisfying
\begin{displaymath}
\inner{\vect{u}_{1}}{\vect{u}_{2}} \approx \tilde{\rho}
\norm{\vect{u}_{1}} \norm{\vect{u}_{2}},
\end{displaymath}
the codeword pair $(\hat{\vect{u}}_{1}, \hat{\vect{u}}_{2}) \in
\mathcal{C}_1 \times \mathcal{C}_2$ whose weighted sum $\alpha_1
\hat{\vect{u}}_{1} + \alpha_2 \hat{\vect{u}}_{2}$ has the smallest
angle to the channel output $\vect{y}$, i.e.
\begin{displaymath}
(\hat{\vect{u}}_{1}, \hat{\vect{u}}_{2}) =
\argmax_{\stackrel{(\vect{u}_1,\vect{u}_2) \in \mathcal{C}_1 \times
  \mathcal{C}_2:}{\inner{\vect{u}_{1}}{\vect{u}_{2}} \approx
  \tilde{\rho} \norm{\vect{u}_{1}} \norm{\vect{u}_{2}}}}
\inner{\frac{\alpha_1 \vect{u}_1 + \alpha_2 \vect{u}_2}{\norm{\alpha_1
      \vect{u}_1 + \alpha_2
      \vect{u}_2}}}{\frac{\vect{y}}{\norm{\vect{y}}}}. 
\end{displaymath}
The corresponding source estimates are then 
\begin{align*}
\hat{\vect{s}}_1 &= \beta_1 \hat{\vect{u}}_1 + \gamma_1 \hat{\vect{u}}_2\\
\hat{\vect{s}}_2 &= \beta_2 \hat{\vect{u}}_1 + \gamma_2 \hat{\vect{u}}_2,
\end{align*}
where the coefficients $\beta_1$, $\gamma_1$, $\beta_2$, $\gamma_2$
are chosen such that $(\hat{\vect{s}}_1, \hat{\vect{s}}_2)$ would form
the minimum mean squared-error estimates of $(\vect{s}_1, \vect{s}_2)$
if $S_1,S_2,U_1^{\ast},U_2^{\ast}$ were zero-mean joint Gaussians with
% covariance
% \begin{displaymath}
% \cov{} = \left( \begin{array} {c c c c}
% \sigma^2 & \rho \sigma^2 & \sigma^2(1-2^{-2R_1}) & \rho
% \sigma^2(1-2^{-2R_2})\\
% \rho \sigma^2 & \sigma^2 & \rho \sigma^2(1-2^{-2R_1}) & \sigma^2 (1-2^{-2R_2})\\
% \sigma^2(1-2^{-2R_1}) & \rho \sigma^2(1-2^{-2R_1}) & \sigma^2
% (1-2^{-2R_1}) & \frac{\tilde{\rho}^2 \sigma^2}{\rho}\\
% \rho \sigma^2(1-2^{-2R_2}) & \sigma^2 (1-2^{-2R_2}) &
% \frac{\tilde{\rho}^2 \sigma^2}{\rho} & \sigma^2 (1-2^{-2R_2})
% \end{array} \right).
% \end{displaymath}
correlation coefficients
\begin{displaymath}
\begin{array} {l l}
\rho(S_1,S_2) = \rho, & \rho(S_1,U_1^{\ast}) = \sqrt{1-2^{-2R_1}} \\[3mm]
\rho(S_1,U_2^{\ast}) = \rho\sqrt{1-2^{-2R_2}}, & \rho(S_2,U_1^{\ast}) =
\rho\sqrt{1-2^{-2R_1}} \\[3mm]
\rho(S_2,U_2^{\ast}) = \sqrt{1-2^{-2R_2}}, &
\rho(U_1^{\ast},U_2^{\ast}) = \tilde{\rho}.
\end{array}
\end{displaymath}

The analysis of the three error events $\left\{ \hat{\vect{u}}_{1}
  \neq \vect{u}_1^{\ast}, \hat{\vect{u}}_{2} = \vect{u}_2^{\ast}
\right\}$, $\left\{ \hat{\vect{u}}_{1} = \vect{u}_1^{\ast},
  \hat{\vect{u}}_{2} \neq \vect{u}_2^{\ast} \right\}$, and
$\left\{ \hat{\vect{u}}_{1} \neq \vect{u}_1^{\ast}, \hat{\vect{u}}_{2}
  \neq \vect{u}_2^{\ast} \right\}$
gives that reliable transmission of the pair $(\vect{u}_1^{\ast},
\vect{u}_2^{\ast})$ is possible for all rates $(R_1,R_2)$ in the
region\footnote{These rate constraints are similar to Ozarow's 
  capacity result for the Gaussian multiple-access channel with
  feedback \cite{ozarow84}.}
\begin{align*}
\mathcal{R} &= \Bigg\{ (R_1,R_2): 
R_1 < \frac{1}{2} \log_2 \left(
  \frac{P_1(1-\tilde{\rho}^2)+N}{N(1-\tilde{\rho}^2)} \right)\\[3mm]
&\qquad \qquad \: \: \: R_2 < \frac{1}{2} \log_2 \left(
  \frac{P_2(1-\tilde{\rho}^2)+N}{N(1-\tilde{\rho}^2)} \right)\\[3mm]
&\qquad \; R_1 + R_2 < \frac{1}{2} \log_2 \left( \frac{P_1+P_2
    +2\tilde{\rho}\sqrt{P_1P_2}+N}{N(1-\tilde{\rho}^2)} \right) \Bigg\}.
\end{align*}
It can then be shown that for all $(R_1,R_2) \in \mathcal{R}$, the
proposed sequence of schemes achieves the distortions\footnote{These
  expressions are similar to the single-rate constraints in the
  quadratic Gaussian two-terminal source coding result
  \cite{wagner_tavildar_vishwanath05, oohama97}.}
\begin{align*}
D_1 &= \sigma^2 2^{-2R_1} \cdot \frac{1-
  \rho^2(1-2^{-2R_2})}{1-\tilde{\rho}^2}\\
D_2 &= \sigma^2 2^{-2R_2} \cdot \frac{1-
  \rho^2(1-2^{-2R_1})}{1-\tilde{\rho}^2}.
\end{align*}

\end{document}